\documentclass[twocolumn,preprintnumbers,amssymb,pra]{revtex4}
\usepackage{amssymb}
\usepackage{url}
\usepackage{ragged2e}
\usepackage[breaklinks,colorlinks,citecolor=blue,urlcolor=violet]{hyperref}
\usepackage{amsmath}
\usepackage{graphicx}
\usepackage{dcolumn}
\usepackage{bm}
\usepackage{verbatim}
\usepackage{setspace}
\usepackage{array}
\usepackage{epsfig}
\usepackage{amsbsy}
\usepackage{xcolor}
\usepackage{soul}
\usepackage{longtable}
\usepackage{enumitem}
\usepackage{natbib}
\usepackage[mediumqspace,mediumspace,amssymb,Gray]{SIunits}

\usepackage{braket}

\newcommand{\threej}[6]{\ensuremath{\left({#1\atop #4}{#2\atop #5}
{#3\atop #6}\right)}}
\newcommand{\sixj}[6]{\ensuremath{\left\{{#1\atop #4}{#2\atop #5}
{#3\atop #6}\right\}}}

\begin{document}
\preprint{\hfill\parbox[b]{0.3\hsize}{ }}


\title{Stabilization method with Relativistic Configuration-interaction  applied to two-electron resonances }
\author{P. Amaro$^{1}$,
        J. P. Santos$^{1,}$\footnote{Email: \url{jps@fct.unl.pt}},
        S. Bhattacharyya$^{2}$, 
        T. K. Mukherjee$^{3}$, 
        J. K. Saha$^{4,}$\footnote{Email: \url{jsaha84@gmail.com}}
          }
\affiliation
{\it
$^{1}${Laboratory for Instrumentation, Biomedical Engineering and Radiation Physics~(LIBPhys-UNL)}, \\Department of Physics, NOVA School of Science and
Technology,~FCT,~NOVA University Lisbon, 2829-516 Caparica, Portugal\\
$^{2}$Department of Physics, Acharya Prafulla Chandra College, New Barrackpore, Kolkata 700131, India\\
$^{3}$Department of Physics, Narula Institute of Technology, Agarpara, Kolkata 700109, India\\
$^{4}$Department of Physics, Aliah University, IIA/27, Newtown, Kolkata 700160, India\\
}

\date{\today}
\date{Received: \today  }


\begin{abstract}
We applied a relativistic configuration-interaction (CI) framework  to the  stabilization method  as an approach  for obtaining the autoionization resonance structure of heliumlike ions. In this method, the ion is confined within an impenetrable spherical cavity, the size of which determines the radial space available for electron wavefunctions and electron-electron interactions. By  varying the size of the cavity, one can obtain the autoinization resonance position and width. The applicability of this method is tested on the resonances of He atom while comparing with benchmark data available in the literature. The present method is further applied on the determination of the resonance structure of heliumlike uranium ion, where a relativistic framework is mandatory. In the strong-confinement region, the present method can be useful to simulate the properties of an atom or ion under extreme pressure.  An exemplary application of the present method to determine the structure of ions embedded in dense plasma environment is briefly discussed.
\end{abstract}


\maketitle

\section{Introduction}
Resonance states are formed in different types of scattering experiments and the nature of interaction with the continuum of the target quantifies the width (or lifetime) of a resonance state. Due to the advancement of experimental techniques and with the advent of free electron lasers, recent interest has been generated to study the resonances of multiply-charged ions \cite{Rudek_2012, Tejaswi_2020, Barmaki_2020, dutta2019, Nrisimhamurty2015}. Different theoretical approaches \emph{e.g.} complex coordinate rotation method \cite{HO19831}, Feshbach projection operator method, optical potential method, R-Matrix method \textit{etc.} (see e.g. Ref. \cite{Schneider_2016} and references therein) have been proposed to understand the internal structure of resonances. All these techniques  need either the complete form of the wavefunctions, the use of complex analytic continuation, or the asymptotic form of the wavefunctions. Hazi and co-workers \cite{hazi1970,fels1971} were the pioneers to introduce stabilization method for analyzing resonance states arising from the elastic scattering with a one-dimensional model barrier potential. After the trial of several variants of this method over more than two decades, finally Mandelshtam \emph{et al.} \cite{mandel1993} put forward an elegant method of calculating the width of resonance states from the spectral density of states (DOS) in the neighborhood of resonance position. The idea of this method \cite{mandel1993} is to diagonalize the Hamiltonian of a quantum system with suitable square-integrable real wavefunctions within a box, and investigate the continuum, bound and resonances (autoionizing) states under variations of the box size.  The continuum states of atomic systems are strongly modified by changes of the box size, while in contrast, the resonance states are hardly affected by the variation of box size, if the minimum size of the box is greater than the effective extent of the wavefunction, or the range of the Coulomb potential. For a bound state, the eigenvalue converges to a particular value with the increase of box size \cite{Hylleraas1930,McDonald1933}. The idea of the box is, therefore, to span the radial space in order to distinguish the variations of these three different types of states. There are different adaptions on the variations of this box. The early workers \cite{hazi1970,fels1971,houston1976} defined the box size as a constraint parameter, usually termed as \emph{hard wall}, in a finite basis set method. Later, Ho \cite{ho1979} and  M\"uller \emph{et al.} \cite{Muller1994} modified the idea of \emph{hard wall} by the \emph{soft wall}, which can be considered as an arbitrary real continuous scaling parameter in a finite basis set. This conjecture \cite{Muller1994} proved to be a very successful one to predict parameters for a wide range of resonances of free, confined as well as field induced few-body systems, where Hylleraas type basis sets are used \cite{kar2004,kar2005,kar2005a,kar2006,kar2007,ghoshal2009,saha2009,saha2010, saha2011, saha2013, Saha_2016,Sadhukhan2019, dutta2019}. Accruing larger radial space through the variation of scaling parameter is also computationally more convenient than expanding the basis by including high-lying configurations as the later may introduce the problem of linear dependency for large basis. Stabilization method has further been extended to determine the atomic resonances near metal surfaces \cite{nimb1995}, resonances in molecules \cite{landau2019,GonzalezLezana2002}, core-excited shape resonances in clusters \cite{fennimore2016,Fennimore2018} as well to probe nuclear resonances \cite{Zhang2008}.

In the present work, we propose a hybrid technique where stabilization method has been adopted within the framework of relativistic configuration-interaction (CI) calculation. A correlated two-electron atomic system where the radial space is truncated by an impenetrable spherical box is considered as a bench-test case. By confining a quantum mechanical system within a model impenetrable spherical cavity and imposing appropriate boundary conditions, the basis sets and the matrix elements can be made explicitly dependent on the radius of the sphere. The electron orbitals are obtained from a finite basis set of B-splines defined in this  impenetrable box, thus making this a {\it hard wall} method. This method has a twofold advantage. Firstly, when the size of the box is reasonably small, the effect of confinement modifies the energy levels of the `free' ion and various problems of spatial confinement \textit{e.g.} pressure ionization in plasma environment \cite{salz} can be probed. With the advent of modern high-speed computational resources and experimental techniques for controlling and confining atoms along with their applications in semiconductors and nanotechnology, this topic of confined systems continues to attract interest nowadays. Manifold applications of such confined quantum mechanical systems are available in literature \textit{e.g.} atoms encaged in endohedral fullerenes \cite{Dolmatov2004}, semiconductor quantum dots \cite{Deng1994,Movilla2005,Zhou2012,Saha_2016,past-20}, ion storage \cite{Connerade1997}, warm dense plasmas \cite{saha2015} \textit{etc}. Comprehensive reviews on this topic may be found in Refs. \cite{Jaskolski1996, brandas2009, kdsen2014}. Secondly, ensuring the radius of the cavity large enough, one can employ the stabilization method \cite{Maier_1980} for determining the parameters of resonance states. To the best of our knowledge, none of the existing variants of stabilization method has been applied under relativistic framework for probing the resonances of highly charged ions or in many-body systems. Both the aspects of this spatial confinement within the relativistic stabilization framework  are developed and discussed here. In this `hard wall' approach the radius of the cavity in the relativistic CI method is spanned in order to locate positions and widths of resonances. To validate the method, the resulting positions and widths of low lying resonances of He atom are compared with benchmark non-relativistic calculations \cite{Abrashkevich1992, Burgers_1995}. As example, we consider the case with null total angular momentum. The relativistic stabilization method is further extended for heliumlike uranium ion to determine the resonance parameters. This method can be exploited to determine the structural properties of highly charged ions under spatial confinement and application in dense plasma environment is also briefly presented here. The details of the methodology are given in Sec. \ref{sec:theor} followed by the discussion on the results in Sec. \ref{result}. Final conclusions are given in Sec. \ref{conc}.
%
\section{Theory}
\label{sec:theor}
\subsection{Configuration Interaction}
We followed the configuration-interaction method of Ref.~\cite{Johnson2007}, to solve the two-electron Dirac Hamiltonian, 
\begin{equation}
\hat{H}({\bf r_1},{\bf r_2})= \hat{h}_0({\bf r_1}) + \hat{h}_0({\bf r_2}) + V({\bf r_1},{\bf r_2})
\end{equation}
where $ \hat{h}_0({\bf r})$ is a single-electron Dirac Hamiltonian solved in a B-splines basis set (see Sec.~\ref{finite_basis}) with the nuclear potential being corrected for a nuclear finite-size. 
 The two-electron wavefuction with total angular momentum $J$, and respective projection over $z$ of $M$ and parity $\Pi$, is defined as a linear combination of configuration-state functions (CSF) of these single-electron solutions, given by 
\begin{eqnarray}
&&\Psi_{\Pi}^J=\sum_{x\le y} C_{xy}^{n} \Psi_{xy}^{\mbox{\tiny CSF}}(JM) \nonumber  \\
&=&  \frac{1}{\sqrt{2}}\sum_{x\le y}  \mu_{x y}  C_{xy}  \Braket{ j_x~m_x~j_y~m_y | J~M}  \left|\begin{array}{cc}
\psi_{x}(1) & \psi_{y}(1)\\
\psi_{x}(2) & \psi_{y}(2)\end{array}\right|
. \nonumber\\
 \label{eq:expa_n}
\end{eqnarray}
Here, $ \Psi_{xy}^{\mbox{\tiny CSF}}(JM)$ represents the CSF of two spherically confined hydrogenic orbitals identified by $x$ and $y$ that involves the ground orbitals and excitations from the occupied to virtual orbitals. The term $ \mu_{xy} $ is given by
\begin{equation}
 \mu_{xy} = \left\{ \begin{array}{ll} 1/\sqrt{2}  &  \mbox{if $x=y$}\\
1  &  \mbox{if $x \ne y$} \end{array} \right.,
\label{eq:mu}
\end{equation}
By inserting this two-electron wavefunction in the Dirac equation, the problem of obtaining the respective eigenvalues is equivalent to diagonalizing the following eigenvalue equation
\begin{equation}%
\sum_{x \le y} \left[ (\epsilon_v + \epsilon_w)\delta_{xv}\delta_{yw} + V_{vw, xy} \right] C_{xy} = E C_{vw}~,
\label{eq:CI_eigen}
\end{equation}%
where $\epsilon_{\nu}$ and $E$ represent the hydrogenic energy of an orbital $\nu$ and the total energy, respectively. The matrix element $V_{vw, xy} =\Braket{ \Phi_{vw} (JM) | V | \Phi_{xy}(JM)}$ of the electron-electron Coulomb interaction is given by
\begin{eqnarray}
V_{vw, xy} &=& \mu_{vw} \mu_{xy} \times \nonumber \\ && \sum_k  (-1)^{J+k +j_w + j_x}
\sixj{j_v}{j_w}{J}{ j_y }{ j_x}{ k} T_k(vwxy) \nonumber \\
&&+ (-1)^{k +j_w + j_x}  \sixj{j_v}{j_w}{J}{ j_x }{ j_y}{ k} T_k(vwyx)~.
\end{eqnarray}
The quantity $T_k(vwyx)$ contains  the Coulomb interaction and is given by
\begin{eqnarray}
T_k(vwyx)&=& (-1)^k  \Braket{ \kappa_v || C^k || \kappa_x} \Braket{ \kappa_w || C^k || \kappa_y} \times \nonumber \\ &&R_k(vwyx)~,
\end{eqnarray}
where the reduced matrix element $\Braket{ \kappa_v || C^k || \kappa_x}$ evaluates to
\begin{eqnarray}
\Braket{ \kappa_v || C^k || \kappa_x}&=& (-1)^{j_v +1/2}\sqrt{[j_v, j_x]}  \times \nonumber \\  &&\threej{j_v}{j_x}{ k}{-\frac{1}{2}}{\frac{1}{2}}{0} \Pi( l_v+k+l_x) ~,
\end{eqnarray}
with  $\Pi(l)$ being the parity term. 
%
%
%
%
The Slater integrals $R_k(vwxy)$, defined by
\begin{eqnarray}\label{Rk}
&&R_k(vwxy) =\\
&&\int_0^{R}dr_1 \left[ P_v(r_1) P_x(r_1) + Q_v(r_1) Q_x(r_1) \right]  v_k(w, y, r_1), \nonumber
\label{eq:R_k}
\end{eqnarray}
contains the overlaps of the large $P(r)$ and small $Q(r)$ components of the hydrogenic orbitals, 
\begin{eqnarray}\label{vk}
&&v_k(w,y, r_1)= \\
&&\int_0^{R} d r_2 \frac{r_{<}^k}{r_{>}^{k+1}}\left[ P_w(r_2) P_y(r_2) + Q_w(r_2) Q_y(r_2) \right]~.   \nonumber
\label{eq:v_k}
\end{eqnarray}
%
Here, $R$ is the radius of the cavity. Although we did not include Breit interaction in this calculation, as shall see in Sec.~\ref{nume_stabi}, differences in energies with {\it state-of-the-art} calculations are of 0.2-0.3\% in heliumlike uranium.  Therefore, although Breit interaction is important for precise determination of the resonances and critical radius (see Sec.~\ref{result}) beyond this uncertainty, the inclusion of only Coulomb potential is (nevertheless) sufficient as {\it proof-of-principle} of the CI to the stabilization method.

We now describe the method of the finite basis set with B-splines for which we obtain the hydrogenic wavefunctions confined in a cavity.
\subsection{Finite basis set with B-splines}
\label{finite_basis}

In the finite basis set approach, the atomic or molecular system is enclosed in a finite cavity with a radius $R$. This leads to a discretization of the continua and, hence, to a representation of the entire Dirac spectrum
in terms of the pseudo-state basis functions. A (quasi--complete) finite set of these states are determined subsequently by
making use of the variational Galerkin method~\cite{Johnson1988}.
In this method, the  action $S_{\kappa}$ is defined as

\begin{eqnarray}
  S_{\kappa} &=&  \frac{1}{2} \int_{0}^{R}\left\{ c P_{n\kappa }(r)
O_{-}^{\kappa}
Q_{n\kappa }(r)\right.
  -c Q_{n\kappa }(r)
O_{+}^{\kappa}
P_{n\kappa }(r)  \nonumber \\
 & & + V(r)\left[ P_{n\kappa }(r)^{2}+Q_{n\kappa
    }(r)^{2}\right]
   \left. -2m_{e}c^{2}Q_{n\kappa }(r)^{2}\right\} dr  \nonumber \\
& & -\frac{1}{2} \, \epsilon \,
  \int_{0}^{R}\left[ P_{n\kappa }(r)^{2}+Q_{n\kappa }(r)^{2}\right] dr
 +  S_{\kappa}^{\mathrm{bond}}
  ~,
  \label{rde1}
\end{eqnarray}
%
from which the Dirac equation states can be derived from the least action principle, $ \delta S_{\kappa}=0$.  To shorten the expressions, we introduced here the operator $O_{\pm}^{\kappa} = \frac{d}{dr} \pm \frac{\kappa }{r}$.
%
%

%
The parameter $\epsilon $ is a Lagrange multiplier introduced to ensure the normalization constraint (Eq.~(\ref{rde1})).
Here, the large, $P_{n\kappa }(r)$, and small, $Q_{n\kappa }(r)$, radial components of the electron wavefunctions can be
written as a finite expansion
\begin{eqnarray}
\begin{array}{ll}
 P(r) =\displaystyle{\sum_{i=1}^{N}} p_{i}B_{i}(r),  & Q(r) =\displaystyle{\sum_{i=1}^{N}} q_{i}B_{i}(r),
 \end{array}
  \label{rde4}
\end{eqnarray}
over the B-splines basis set $B_{i}(r)$ that are given in detail in subsection \ref{sec:bspl}. 
Moreover, in Eq.~(\ref{rde4}), the subscripts $n$ and $\kappa $ have  been omitted from the functions $P_{n\kappa}(r)$ and $Q_{n\kappa}(r)$ for the sake of notation simplicity.
The function $S_{\kappa}^
{\mathrm{Bond}}$ in Eq.~(\ref{rde1}), given by
\begin{eqnarray}
  &&S_{\kappa}^{\mathrm{Bond}} = \\
  &&\left\{
    \begin{array}{l}
      \frac{c}{4} \left[ P^{2}(R) -Q^{2} (R) \right] +\frac{c}{2} P(0)
      \left[ P(0) - Q(0) \right]  \\ \hspace{6cm}~ \rm{for  }~  \kappa<0 \\
      \frac{c}{4} \left[ P^{2}(R) -Q^{2} (R) \right] +\frac{c}{2} P(0)
      \left[2 c P(0) - Q(0) \right] \\ \hspace{6cm}~ \rm{for  }~ \kappa>0 \\
    \end{array}
  \right. ~,  \nonumber
\label{SBound_01}
\end{eqnarray}
 assures the boundary conditions $P(0)=0 \hspace{0.5cm} \mathrm{and} \hspace{0.5cm} P(R)=Q(R)$ known as MIT-bag-model condition \cite{Chodos1974}, was included to avoid the Klein's paradox, which arises when one attempts to confine a particle to a cavity, essentially by forcing the radial current crossing the boundary to vanish \cite{Greiner1990}. 
 
Inserting the radial components (\ref{rde4}) into the least action principle (\ref{rde1}) and evaluating the variation $S_{\kappa}$ with respect to change of the expansion coefficients $p_{i}$ and $q_{i}$, we obtain the matrix equation
\begin{equation}
  A v = \epsilon B v  ~,
  \label{rde5}
\end{equation}
where $ v= \left(p_{1},p_{2},\ldots ,p_{N}, q_{1},q_{2},\ldots ,q_{N} \right)$. $A$ and $B$ are symmetric $2N \times 2N$
matrices given respectively by
%
\begin{equation}
  A=
  \left[
    \begin{array}{cc}
      (V) & c\left[ (D)- \left( \displaystyle\frac{\kappa }{r} \right)
      \right]  \\
      -c\left[ (D)+ \left( \displaystyle\frac{\kappa }{r} \right)
      \right]  & -2 c^{2}(C)+(V)
    \end{array}
  \right]
  +  A^{\mathrm{bond}}~,
  \label{rde7}
\end{equation}
%
and
\begin{equation}
  B=
  \left[
    \begin{array}{cc}
      (C) & 0 \\
      0 & (C)
    \end{array}
  \right] .
  \label{rde8}
\end{equation}
The matrix $A^{\mathrm{bond}}$ reflects the boundary conditions, and is defined by
\begin{equation}
A_{ij}^{\mathrm{bond} }=\left\{
\begin{array}{ll}
c\delta _{i,1}\delta _{j,1}-\frac{c}{2}\delta _{i,1}\delta _{j,n+1}-\frac{c}
  {2}\delta _{i,n+1}\delta _{j,1}    & \\
  +\frac{c}{2}\delta _{i,n}\delta _{j,n}-\frac{c}{2}\delta _{i,2n}\delta_{j,2n} & \rm{if}~ \kappa \leq 0 , \\
2c^{2}\delta _{i,1}\delta _{j,1}-\frac{c}{2}\delta _{i,1}\delta _{j,n+1}-%
\frac{c}{2}\delta _{i,n+1}\delta _{j,1} & \\
+\frac{c}{2}\delta _{i,n}\delta _{j,n}-\frac{c}{2}\delta _{i,2n}\delta
_{j,2n}& \rm{if}~ \kappa >0 ~.%
\end{array}%
\right.  \label{matr_A_li}
\end{equation}
The $N\times N$ matrices $(C)$, $(D)$, $(V)$ and $
(\kappa /r)$ are given by
\begin{eqnarray}
  (C)_{ij}&=&\int B_{i}(r)B_{j}(r)dr ~,
  \label{rde10a}
 \\
  (D)_{ij}&=&\int B_{i}(r)\frac{d}{dr} B_{j}(r) dr,
  \label{rde10b}
 \\
  \left( \frac{\kappa}{r} \right)_{ij}&=&\int B_{i}(r) \frac{\kappa}{r} B_{j}(r)dr.
  \label{rde10c}
 \\
  (V)_{ij}&=&\int B_{i}(r) V(r) B_{j}(r) dr \, .
  \label{rde10d}
\end{eqnarray}
%


Equation (\ref {rde5}) is known as a generalized eigenvalue problem that can be solved by employing the standard
techniques from the linear algebra. 
In the present work, we have used the well-established Linear Algebra PACKage 3.3.1 (LAPACK\label{lab_lapa}) \cite{lapack1999}.

\subsection{B-splines}
\label{sec:bspl}

Following the de Boor textbook \cite{Boor1978}, we divide the interval of interest $[0, R]$
into segments whose endpoints define a knot sequence $\{t_{i}\}=1,2,\ldots , N+k.$, where $R$ is the cavity radius.
The B--splines of the order $k$, $B_{i,k}(r)$, are defined on this knot sequence by the recurrence relation
\begin{equation}
  B_{i,k}(r) = \frac{r-t_{i}}{t_{i+k-1}-t_{i}}B_{i,k-1}(r)
  + \frac{t_{i+k}-r}{t_{i+k}-t_{i+1}}B_{i+1,k-1}(r) ,
\label{bs2}
\end{equation}
where the B--splines of the first order read as
\begin{equation}
  B_{i,1}(r)=\left\{
  \begin{array}{l}
  1,\qquad t_{i}\leq r\leq t_{i+1} \\
  0,\qquad \mbox{otherwise}
\end{array}
\right. \, .
\label{bs1}
\end{equation}
%
%
The first 1, 2... $k$ and the last $N+1$, $N+2$... $N+k$ knots must be equal and are defined as: $t_{1} = t_{2} = \ldots = t_{k} = 0$ and $t_{N+1} = t_{N+2} = \ldots = t_{N+k} = R$. Otherwise, the $t_{i}$ knots with $k<i<n$ follow an exponential grind.

 Eigenstates of Eq.~\eqref{rde5} labeled by $i=1, ... ,N$ addresses the negative continuum $\epsilon_i<-2mc^2$ while solutions with  $i=N +1, ... ,~2N$ describe bound states (first few ones) and the continuum $\epsilon_i>0$.

\subsection{Numerical stability of the CI method}
\label{nume_stabi}

Before providing results for the application of the CI method to the stabilization method it is worthwhile to attest the quality of the obtained states. Following previous works with single-electron spectrum constructed from B-splines (e.g. \cite{Santos1998, Amaro2009,  Safari2012, Amaro2016}), the spectrum was attested  with variations of the B-splines parameters, namely the order ($k$) and number of splines ($N$). Optimal values of these parameters are $k=9$ and $N>60$, matching with analytical solutions at 13 digits for the first five bound states. This optimization was made for a large enough radius of $R=40$, in which cavity effects are not present. Through all calculations, these parameters were set constant with exception of the radius. The effect of finite-nuclear size was found negligible within the quoted precision for all values obtained from the stabilization method presented in Sec.~\ref{result}.  

These  single-state orbitals were included in the CI matrix \eqref{eq:CI_eigen} for diagonalization. Since the finite-basis method discretize the continua, spurious autoionizing states can result from  this discretization of the negative and positive continua. In detail, this  could led to a sum of discrete negative eigenstates and positive ones in the CI diagonal terms, which could have energies close to two-electron bound states, and thus distort the spectrum. To  avoid these spurious autoionizing states, discrete negative states were not included in the CI matrix. Excluding these states is similar to the projection method in multiconfiguration Dirac-Fock method \cite{Indelicato1995a}.  For the case of $\Pi=0$ and $J=0$ spectrum, orbitals of type $nsn's$, $npn'p$, $nsn'd$ and $ndn'd$, with $n\le3$ and $n'\le10$, were included to form two-electron wavefunction \eqref{eq:expa_n}. Hereafter, we refer each eigenvalue by the configuration of the dominant CSF, e.g., the ground state with $\Pi=0$ and $J=0$ is referred as $1s^2$. Further increase of these maximum number of orbitals and orbital momentum converge to changes in the 6 digit. Table~\ref{tab:energ} contains  the ground state $1s^2$ and first two excited states \textit{i.e.} $1s2s$ and $1s3s$ calculated for $Z=92$, as well as {\it state-of-the-art} calculations in Ref.~\cite{Yerokhin2019} that contains full inclusion of Breit interaction and QED effects. The energy values are in  agreement with this reference within 0.2-0.3\% relative difference, which is attributed to these extra QED effects.

All obtained eigenvalues for the application of stabilization method have $\Pi=0$ and $J=0$.
\begin{table}[t]
\caption{\label{tab:energ} Eigenvalues (a.u.) with $\Pi=0$ and $J=0$ for the ground state ($1s^2$) and first two excited states ($1s2s,1s3s$) in helium-like U ($Z=92$). Values are compared with respective ones listed in Ref.~\cite{Yerokhin2019}.
}
\begin{ruledtabular}
\begin{tabular}{llll}
&$E_{1s^2}$ &$E_{1s2s}$ & $E_{1s3s}$ \\ 
\hline
This work & --9636	&	--6081	&	--5380		\\
Ref.\cite{Yerokhin2019} & --9605 & --6067 & --5369
\end{tabular}
\end{ruledtabular}
\end{table}

\section{Results and Discussions}
\label{result}

\begin{figure}[t]
\centering
\includegraphics[clip=true,width=0.95\columnwidth]{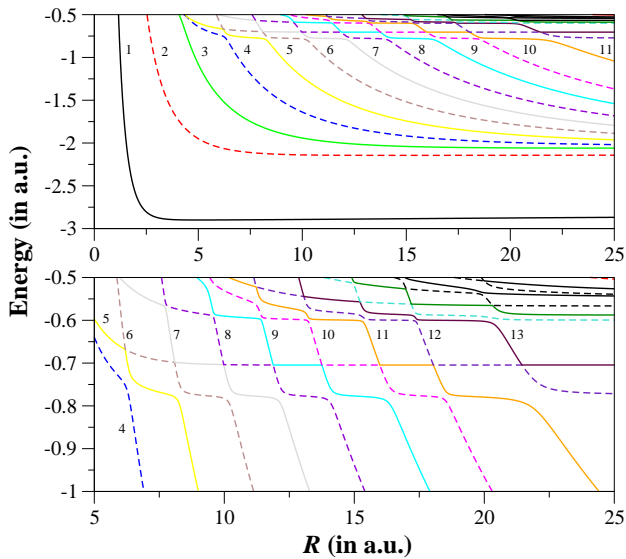}
\caption{ (Color online)  Stabilization diagram constructed with the energy eigenvalues obtained after solving Eq.~\ref{eq:CI_eigen} with $\Pi=0$ and $J=0$ in function of the cavity radius and for helium atom ($Z=2$). Configurations state functions up to $l=2$ are included in the calculation. Each line (with specific color) represents an eigenvalue sorted by energy \textit{e.g.} black line represents the 1st eigenroot, red line represents the 2nd eigenroot, green line represents the 3rd eigenroot and so on. Note that the lower panel uses the same colour code for the energy eigenroots as that of the upper panel.}
\label{fig:stab-he}
\end{figure}

Stabilization diagram constructed with the first $20$ energy eigenvalues of He atom lying between $-3.0$ a.u. and $-0.5$ a.u. is given in the upper panel of Fig.~\ref{fig:stab-he}. The cavity radius $R$ is varied in the range $1.0$ a.u. to $25.0$ a.u. in the mesh size of 0.01 a.u. It is evident from Fig.~\ref{fig:stab-he} that the first energy eigenvalue \textit{i.e.} the ground state ($1s^2$) is almost insensitive \textit{w.r.t.} $R$ $\in$ [2, 25] a.u. Before going into the resonance structure of the system for high $R$, we analyze the features observed in the strong confinement region. Decrease of $R$ below 2 a.u. causes a drastic effect on the ground state energy of He. In fact, the curve exhibits a sharp bend \textit{i.e.} a `knee' and becomes nearly parallel to the vertical (energy) axis leading to fragmentation threshold for $R \sim 1.1$ a.u. For the second and the third energy eigenvalues, \textit{i.e.},  the first $1s2s$ and the second $1s3s$ singly excited states of He, such `knee' occurs at larger values of cavity radius ($R$). The energy eigenvalues of $1sns$ ($n=1-3$) of He atom, for different values of cavity radius ($R$) are given in Table \ref{tab:shifts}. The present energy values are compared with the benchmark non-relativistic results of Bhattacharyya \textit{et al.} \cite{Bhattacharyya_2013} calculated with Hylleraas-type basis set.
 In order to estimate the relativistic contribution, we evaluate these energies with Eqs.~\eqref{eq:R_k} and \eqref{eq:v_k}  having the full relativistic wavefunction (large and small components), and without the small component, respectively.  We note that the present CI method gives lower values of total energy compared with Ref. \cite{Bhattacharyya_2013}, as cavity radius decreases. As observed in Table~\ref{tab:shifts}, this is not due to the inclusion of relativity since an estimation of these effects based on the small component returns a negligible influence.  These differences between relativistic and non-relativistic values are slightly amplified on lower values of $R$ where the total energies are more sensitive to this parameter. A closer look to the CI matrix \eqref{eq:CI_eigen} at $R=1.11$~a.u. and to the first two diagonal terms ($1s^2$ and $1s2s$) shows that the orbital energies $2 \epsilon_{1s}$ are set apart from $ \epsilon_{1s}+\epsilon_{2s}$ by $\sim$10~a.u., while Coulomb repulsion between these states equals $V_{1s^2,1s2s}=0.6$~a.u..  This makes the $1s^2$ state less influenced by mixing of $1s2s$ state, as well as by the rest of the spectrum. As consequence, the  value of $1s^2$ state at $R=1.11$  is mainly given by $2 \epsilon_{1s} + V_{1s^2,1s^2}$.   Further verification from Hylleraas method can thus be traced back to these terms. 
  The atomic electrons cease to be in a bound state if the cavity radius decreases below a critical value, denoted as $R_c$. In case of the $1s^2$ state this occurs when the orbital energies ($\epsilon_{1s}$) matches the Coulomb repulsion energy ($V_{1s^2,1s^2}$). The critical cavity radii ($R_c$) corrected up to two decimal place for $1s^2$, $1s2s$ and $1s3s$ states are estimated as 1.10 a.u., 2.30 a.u. and 3.08 a.u. respectively. The bound electron(s) will detach from the atom if $R < R_c$ and the cavity-bound free electron continuum becomes discrete. But still the electrons in the continuum remain bound by the cavity and entangled to the parent ion \cite{Koscik2015}. We have also calculated these quantized positive energy states of the electrons in presence of the ion in the centre of the confining sphere. For $R = $1.0 a.u., energy of the ground state of cavity bound atom is 0.9769 a.u., the first excited state is 13.8676 a.u. and the second excited state is 14.3220 a.u. Further reduction of $R$ to 0.5 a.u. yields these values at 22.2728 a.u., 67.3502 a.u. and 68.4632 a.u. respectively. The present results are in reasonable agreement with the non-relativistic results \cite{riveros_2010,Koscik2015} estimated in correlated Hylleraas basis. 
  
Under an adiabatic approximation, the amount of pressure ($P$) `felt' by the system inside the impenetrable cavity can be expressed as
\begin{eqnarray}\label{pres}
P=-\frac{1}{4\pi R^2}\frac{dE_g}{dR}\simeq -\frac{1}{4\pi R^2}\frac{\Delta E_g}{\Delta R}~,
\end{eqnarray} 
where $E_g$ is the ground state energy of the system inside the sphere of radius $R$. For higher excited states, this relation may not be applicable as the equilibrium criteria is not satisfied because of finite lifetimes of such states. However, we can assume that the amount of pressure experienced by the ion
would be same for all the states at a particular value of $R$. Therefore, the pressure ($P$) at a particular value of $R$ felt by an ion in an excited state can be estimated from Equation \ref{pres} by calculating the energy gradient of the ground state $1s^2$ \textit{w.r.t.} $R$ around the same value of $R$. In the present calculation, we have taken $\Delta R =10^{-4}$ a.u. The critical pressure ($P_c$) or the ionization pressure \textit{i.e.} the pressure experienced by the He atom at the critical cavity radius ($R_c$) values 1.10 a.u., 2.30 a.u. and 3.08 a.u. are estimated as 1611 GPa, 150.4 GPa and 14.4 GPa respectively. These $P_c$ values are in excellent agreement with that of Saha \textit{et. al.} \cite{saha2016} and establishes the applicability of the present method in strong confinement region.

\begin{table}[t]
\caption{\label{tab:shifts} First three eigenvalues ($1sns$ [$n=1-3$])  of helium atom for several confinement radii $R$ (a.u.) calculated in the present work. $E_{\mbox{\tiny L+S}}$ and  $E_{\mbox{\tiny L}}$ and are the energies calculated with the CI method having the large and small components of the wavefunctions, and only the large component, respectively. Non-relativistic results of Bhattacharyya \textit{et al.} \cite{Bhattacharyya_2013} are shown for comparison.
}
\begin{ruledtabular}
\begin{tabular}{llll}
$R$(a.u.) & $E_{\mbox{\tiny L+S}}$ &  $E_{\mbox{\tiny L}}$ & Ref.~\cite{Bhattacharyya_2013}\\
\hline
1.11&	--0.0988	&	--0.0991	&	--0.0739	\\
1.15&	--0.4004	&	--0.4007	&	--0.3792	\\
1.3	&	--1.2426	&	--1.2429	&	--1.2310	\\
1.5	&	--1.9116	&	--1.9118	&	--1.9070	\\
2	&	--2.6027	&	--2.6028	&	--2.6040	\\
3	&	--2.8694	&	--2.8695	&	--2.8724	\\
4	&	--2.8973	&	--2.8974	&	--2.9005	\\
5	&	--2.9001	&	--2.9001	&	--2.9034	\\
$\infty$	&	--2.9001	&	--2.9001	&	--2.9037	\\
\hline							
2.31	&	--0.0339	&	--0.0341	&	--0.0283	\\
2.35	&	--0.1282	&	--0.1284	&	--0.1230	\\
2.4	    &	--0.2386	&	--0.2387	&	--0.2339	\\
2.5	    &	--0.4369	&	--0.4371	&	--0.4332	\\
2.7	    &	--0.7607	&	--0.7609	&	--0.7583	\\
3.0	    &	--1.1152	&	--1.1153	&	--1.1141	\\
5.0	    &	--1.9494	&	--1.9494	&	--1.9497	\\
$\infty$	&	--2.1402	&	--2.1401	&	--2.1459	\\
\hline							
3.21	&	--0.0921	&	--0.0921	&	--0.0062	\\
3.3	&	--0.1471	&	--0.1472	&	--0.0724	\\
3.5	&	--0.2507	&	--0.2507	&	--0.2016	\\
4	&	--0.4562	&	--0.4563	&	--0.4582	\\
5	&	--1.0796	&	--1.0797	&	--1.0792	\\
$\infty$	&	--2.0595	&	--2.0595	&	--2.0613	\\   	
\end{tabular}
\end{ruledtabular}
\end{table}

\begin{figure}[b]
\centering
\includegraphics[clip=true,width=1.0\columnwidth]{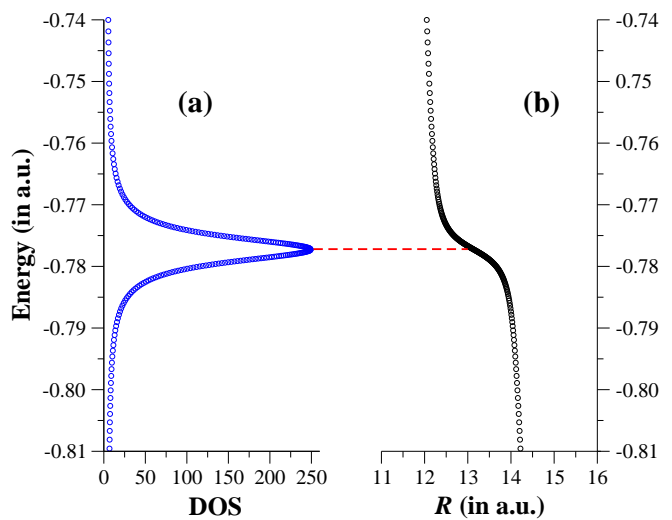}
\caption{ (Color online) Combined plot of numerically estimated DOS (Eq.~\eqref{stbp}) vs. Energy (in a.u.) for eigenvalue no. 8 from the stabilization diagram of Fig.~\ref{fig:stab-he}.}
\label{fig:combo}
\end{figure}

The upper panel of Fig.~\ref{fig:stab-he} reveals that all the higher energy eigenvalues vary significantly \textit{w.r.t.} cavity radius ($R$) and produces a flat, short plateau in the vicinity of avoided crossing. Enlarged view of such avoided crossings within the energy range of -1.0 to -0.5 a.u. is given in the lower panel of Fig.~\ref{fig:stab-he}. It clearly appears that such plateaus occur at some particular energy values that are the positions of the resonances. For instance, it can be observed  that the first resonance position is close to -0.77 a.u. Precise determination of the resonance parameters \textit{i.e.} positions and widths of these states are done in a two-step process. The first step is to take an eigenvalue (single color line in Fig.~\ref{fig:stab-he}) and estimate the spectral density of states (DOS) from the stabilization diagram by taking the inverse of tangent at different points near the stabilization plateau for each energy eigenvalue using the formula:
\begin{eqnarray}\label{stbp}
\rho_{n}(E)=\left|\frac{R_{i+1}-R_{i-1}}{E_{n}(R_{i+1})-E_{n}(R_{i-1})}\right|.
\end{eqnarray}
For a clear visualization, the 8-th energy eigenvalue showing the plateau near -0.77 a.u. and corresponding spectral DOS are plotted within the energy range $-0.81 \leqslant E \leqslant -0.74$ a.u. in Fig.~\ref{fig:combo}. The DOS plot shows a peak in the middle of the plateau (follow the red line) showing the position of the lowest lying ($2s^2$) resonance below He$^+(2s)$.

\begin{figure}[h]
\centering
\includegraphics[clip=true,width=0.95\columnwidth]{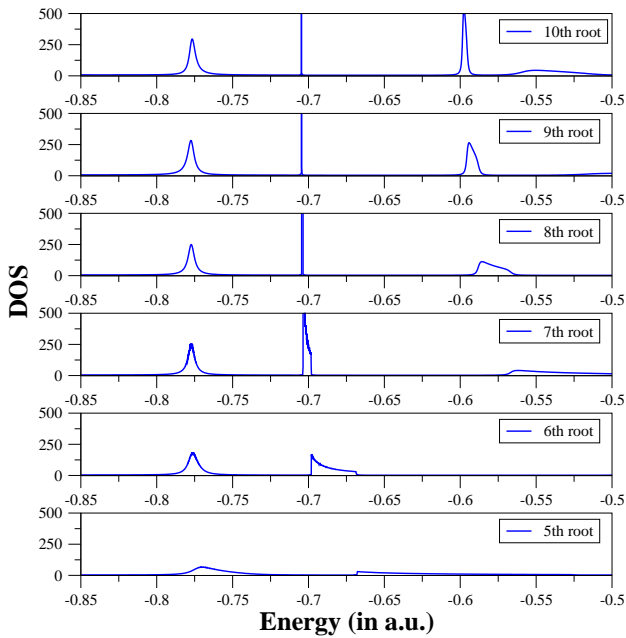}
\caption{ (Color online) The DOS peaks of eigenvalue no. 5 to 10 in the energy range --0.85 a.u. to --0.50 a.u.}
\label{fig:dos-he}
\end{figure}
As one eigenvalue may produce plateaus at different energies (see Fig.~\ref{fig:stab-he}), corresponding peaks of DOS will occur at those values. The DOS profiles of eigenvalues $5$ to $10$ within the range $-0.85 \leqslant E \leqslant -0.50$ a.u. are given explicitly in Fig.~\ref{fig:dos-he}. It is clear that three peaks at three different energies are converging for first three resonances. Among these three DOS peaks, the middle one near $-$0.7 a.u. represents a very narrow width, evident from the plots of $8-10$ eigenvalues. Thus the lifetime of this state is considerably high compared to the other two states lying on either sides. This state corresponds to $2p^2$. The next peak around -0.59 a.u. is the $2s3s$ state while the first peak around -0.77 a.u. corresponds to the $2s^2$ state. It is also evident that these resonances are isolated as the separation of peaks are greater than the widths of the consecutive resonances.

\begin{figure}[t]
\centering
\includegraphics[clip=true,width=0.96\columnwidth]{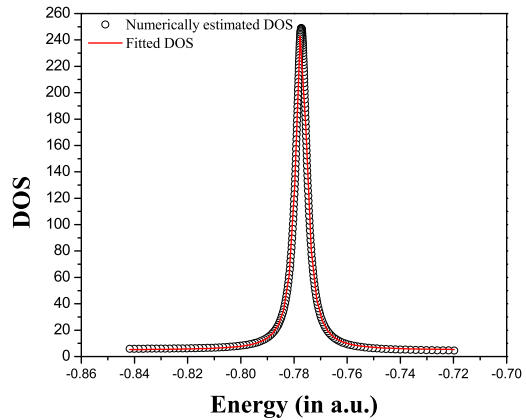}
\caption{ (Color online) Numerically estimated DOS vs. Energy (in a.u.) (black hollow circles) and the fitted Lorentzian  (red line) for eigenvalue no. 8 showing 1st  resonance ($\sim$ --0.77 a.u.) of He below $\mathrm{He^{+}}$($2s$) threshold.}
\label{fig:lorentz-he}
\end{figure}

In the next step, we consider DOS of each isolated resonance and fit it with a Lorentzian profile
\begin{eqnarray}\label{lf}
\rho_{n}(E)=y_{0}+\frac{A}{\pi}\frac{\Gamma/2}{(E-E_{r})^{2}+(\Gamma/2)^{2}}
\end{eqnarray}
where, $\Gamma$ is the full width at half maximum of the peak, $y_{0}$ is the offset, $A$ represents the area under the curve from the base line, and $E_{r}$ is the energy corresponding to maximum of $\rho_{n}$, \textit{i.e.} the resonance position. As an example, the estimated DOS and the fitted Lorentzian  corresponding to eigenvalue no. 8 is given in Fig.~\ref{fig:lorentz-he}. The fitting to this curve [Eq. \eqref{lf}] yields the resonance parameters $E_{r} = -0.77726$ a.u. and $\Gamma = 0.00502$ a.u. ($\chi^2=0.09964, ~\mathcal{R}^2=0.99999$). This is the lowest lying $^1S_0$ resonance below He$^+(2s)$. Repeated calculations of DOS near the plateau of each of the eigenvalues for the resonance states are performed, which result into fitted Lorentzian  similar to Fig.~\ref{fig:lorentz-he}. For a particular resonance, the position and width is chosen with respect to the best fitting parameters. For instance, among the fitting parameters ($\chi^2$, $\mathcal{R}^2$) of the first DOS peak of eigenvalues $8-10$, we find eigenvalue no. 8 yields the least $\chi^2$ value and also $\mathcal{R}^2$ value closer to unity. Similar fitting for the second resonance yields $E_r = -0.70466$ a.u., $\Gamma = 0.0000368$ a.u. while for the third resonance, we find $E_r = -0.59731$ a.u., $\Gamma = 0.0031956$ a.u. The resonance parameters $(E_r, \Gamma)$ for other states can be obtained in a similar manner. The estimated resonance parameters of the first and third resonances of He atom are in good agreement with the benchmark non-relativistic numerical records \cite{Abrashkevich1992, Burgers_1995} for resonances below second ionization threshold of He atom. For instance, Abrashkevich \textit{et. al.} \cite{Abrashkevich1992} reported the resonance positions of first and third resonances as --0.778824 a.u. and --0.590158 a.u.,  respectively by adopting coupled--channel hyperspherical adiabatic approach. Explicitly correlated Hylleraas type basis set in the framework of complex-coordinate-rotation calculation of Burgers \textit{et. al.} \cite{Burgers_1995} yields the highly precise estimate of widths of first and third resonances as 0.004541126 a.u. and 0.001362478 a.u., respectively.

\begin{figure}[b]
\centering
\includegraphics[clip=true,width=0.95\columnwidth]{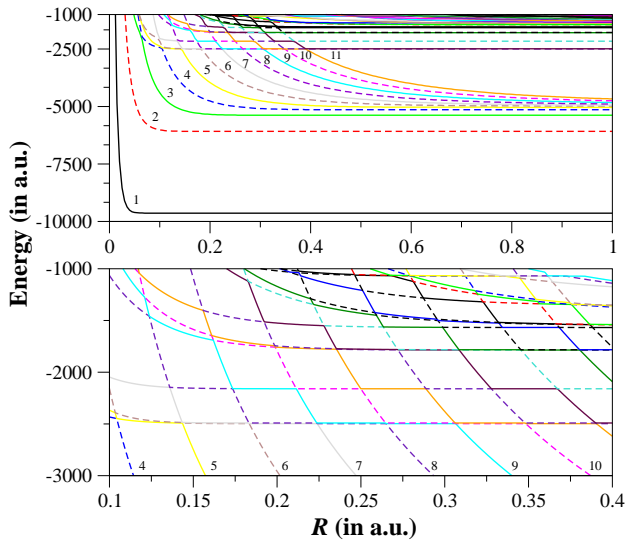}
\caption{ (Color online)   Stabilization diagram constructed with the energy eigenvalues obtained after solving Eq.~\ref{eq:CI_eigen} with $\Pi=0$ and $J=0$ in function of the cavity radius and for heliumlike uranium atom ($Z=92$). Configurations state functions up to $l=2$ are included in the calculation. Each line represents an eigenvalue sorted by energy. Each line (with specific color) represents an eigenvalue sorted by energy \textit{e.g.} black line represents the 1st eigenroot, red line represents the 2nd eigenroot, green line represents the 3rd eigenroot and so on. Note that the lower panel uses the same colour code for the energy eigenroots as that of the upper panel.}
\label{fig:stab-ur1}
\end{figure}

After standardizing the stabilization method within relativistic CI framework for He atom, we extend it to determine the resonance structure of highly charged heliumlike uranium ($Z=92$) where relativistic effects play a major role. The stabilization diagram of $\mathrm{^1S_0}$ states of $\mathrm{U^{90+}}$ is given in Fig.~\ref{fig:stab-ur1}. In the upper panel, the diagram displays the first 20 energy eigenvalues within the range --10000.0 a.u. to --1000 a.u.  The ground ($1s^2$) and first two excited states \textit{i.e.} $1s2s$ and $1s3s$  respectively (energies listed in Table~\ref{tab:energ}), lying below $\mathrm{U^{91+}(1s)}$ threshold, are evident from Fig.~\ref{fig:stab-ur1}.  In the strong confinement regime, these states show the same pattern as of He (Fig.~\ref{fig:stab-he}) except the positions of the respective `knees', which occur at lower values of $R$ as compared to He, due to the contraction of the wavefunctions in the much stronger Coulomb field. 

\begin{figure}[t]
\centering
\includegraphics[clip=true,width=0.95\columnwidth]{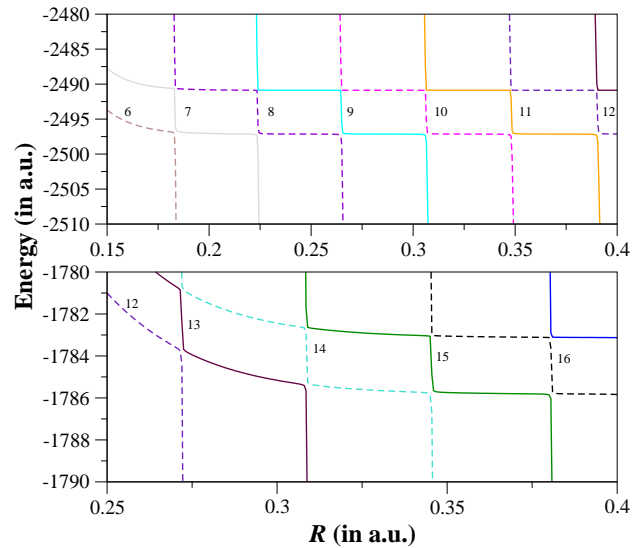}
\caption{ (Color online) Enlarged view of the stabilization diagram given in Fig \ref{fig:stab-ur1}. For both upper and lower panels, the line colors carry similar meaning as of Fig. \ref{fig:stab-ur1}}
\label{fig:stab-ur2}
\end{figure}

\begin{figure}[t]
\centering
\includegraphics[clip=true,width=0.96\columnwidth]{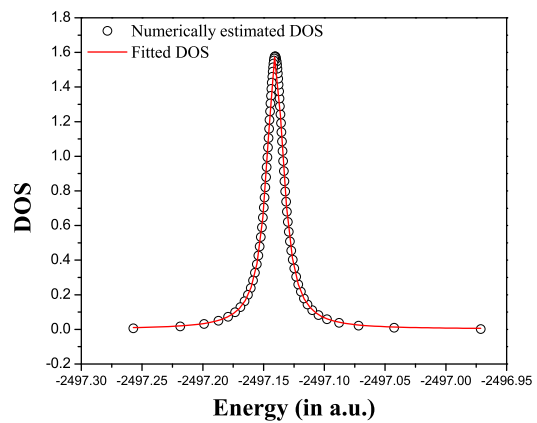}
\caption{ (Color online) Numerically estimated DOS vs. Energy (in a.u.) (black hollow circles) and the fitted Lorentzian  (red line) for eigenvalue no. 11 showing 1st resonance ($\sim$ --2497 a.u.) of heliumlike uranium below $\mathrm{U^{91+}}$($2s$) threshold.}
\label{fig:lorentz-ure1}
\end{figure}

\begin{figure}[t]
\centering
\includegraphics[clip=true,width=0.96\columnwidth]{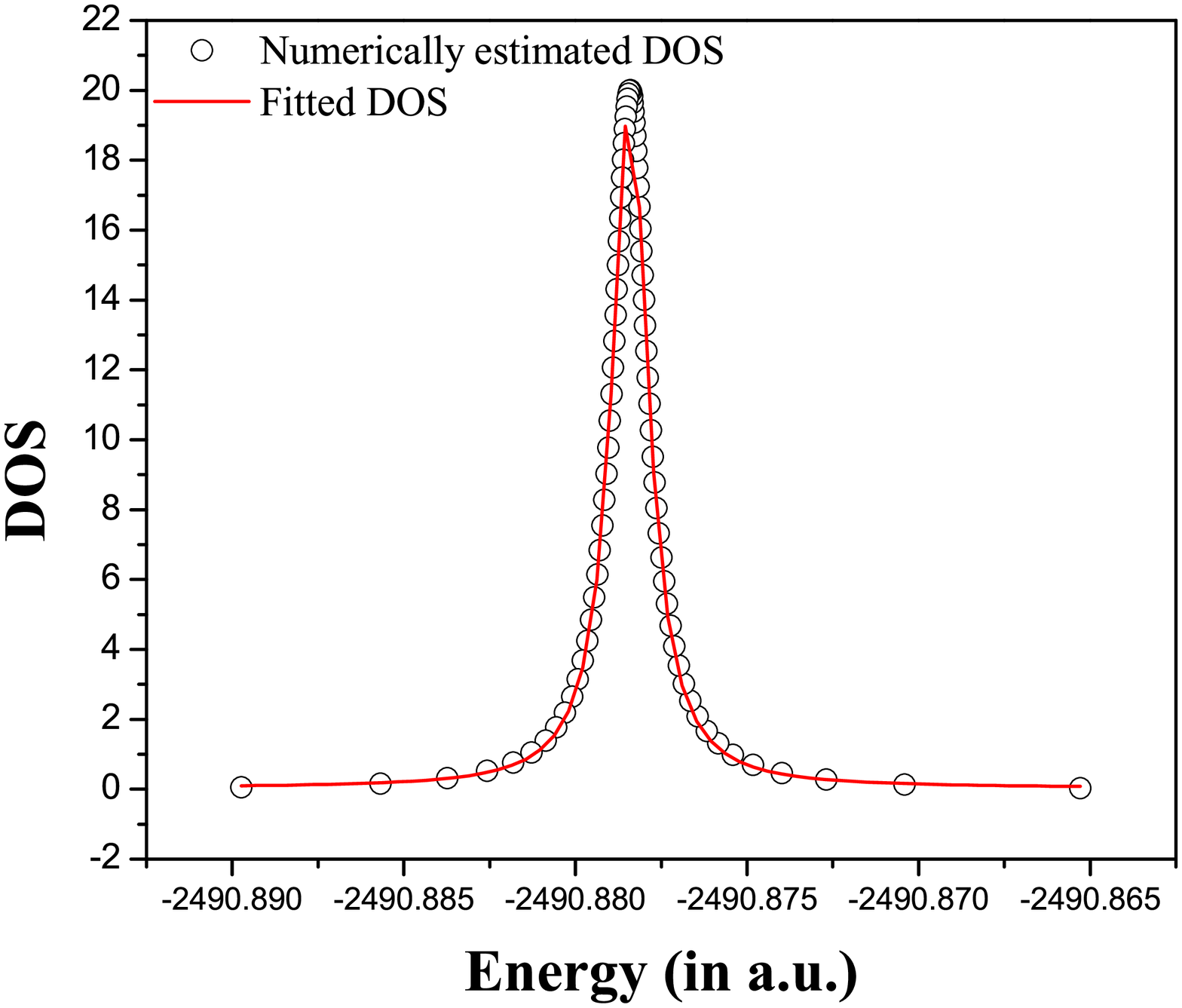}
\caption{ (Color online) Numerically estimated DOS vs. Energy (in a.u.) (black hollow circles) and the fitted Lorentzian  (red line) for eigenvalue no. 11 showing 2nd resonance ($\sim$ --2490 a.u.) of heliumlike uranium below $\mathrm{U^{91+}}$($2s$) threshold.}
\label{fig:lorentz-ure2}
\end{figure}

\begin{figure}[t]
\centering
\includegraphics[clip=true,width=0.96\columnwidth]{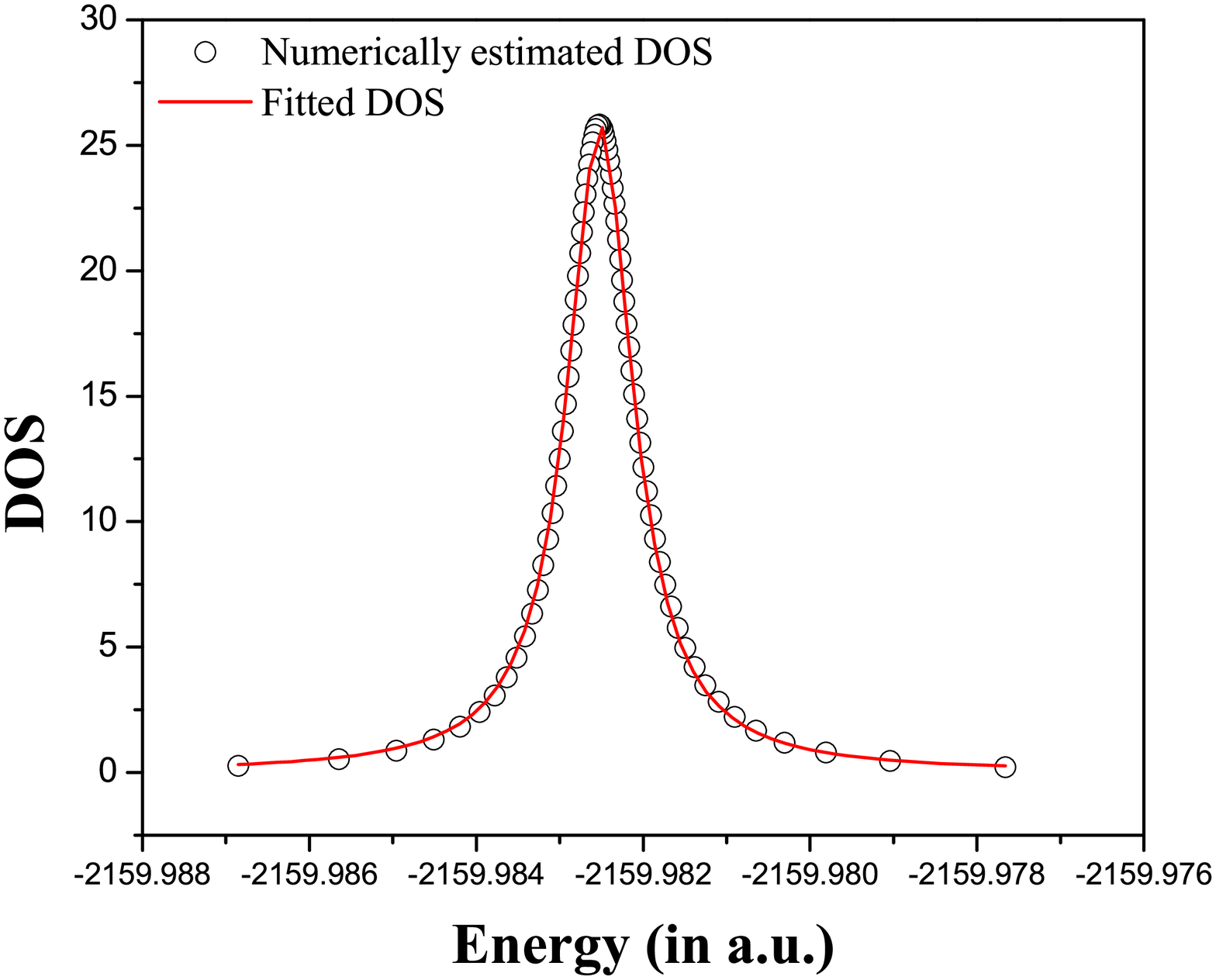}
\caption{ (Color online) Numerically estimated DOS vs. Energy (in a.u.) (black hollow circles) and the fitted Lorentzian  (red line) for eigenvalue no. 12 showing 3rd  resonance ($\sim$ --2159 a.u.) of heliumlike uranium below $\mathrm{U^{91+}}$($2s$) threshold.}
\label{fig:lorentz-ure3}
\end{figure}

An enlarged view of the stabilization diagram of $\mathrm{U^{90+}}$ in the energy range $-3000$ a.u. to $-1000$ a.u. is given in the lower panel of Fig.~\ref{fig:stab-ur1}. The energy range is properly chosen to locate the resonances between $1s$ and $2s$ thresholds of $\mathrm{U^{91+}}$. A series of resonances of heliumlike uranium near the energies --2495 a.u., --2159 a.u., --1783 a.u., --1556 a.u. are seen. A closer look at the avoided crossings of the eigenvalues near the energies --2495 a.u. and --1783 a.u. are shown in the upper and lower panels of Fig.~\ref{fig:stab-ur2} respectively. A pair of close resonances at (--2497.12 a.u., --2490.87 a.u.) in the upper panel and (--1785.57 a.u., --1783.04 a.u.) in the lower panel are visible in Fig.~\ref{fig:stab-ur2}. As the pair of resonances are very closely spaced, we have taken a much lesser mesh size (0.0001 a.u.) of $R$ compared to that taken for He for the precise determination of the DOS profiles. Figure \ref{fig:lorentz-ure1} and \ref{fig:lorentz-ure2} display plot of the numerically estimated DOS vs. energy (in a.u.) for eigenvalue no. 11 showing close lying 1st and 2nd resonances respectively of heliumlike uranium below $\mathrm{U^{91+}}$($2s$) threshold. The fitting to the curve given in Fig. \ref{fig:lorentz-ure1} yields $E_{r} = -2497.1405$~a.u. and $\Gamma = 0.01682$ a.u. ($\chi^2=0.000003, ~\mathcal{R}^2=0.99999$) while the same in Fig.~\ref{fig:lorentz-ure2} yields $E_{r} = -2490.87841$ a.u. and $\Gamma = 0.00126$ a.u. ($\chi^2=0.01625, ~\mathcal{R}^2=0.99998$). The fitting further confirms that the pair of resonances are isolated despite being closely spaced. In a similar fashion using eigenvalue no. 12, we have determined $E_{r} = -2159.9825$ a.u. and $\Gamma = 0.00096$ a.u. for the 3rd resonance and the corresponding plot is given in Fig.~\ref{fig:lorentz-ure3}. The parameters of first six resonance states of heliumlike uranium are depicted in Table \ref{tab:resoure}. Sharp gradual decrement of the width of the resonance states are noticed which may be due to respective increase in the inter-electronic separation. It is to be noted that only the tentative positions are reported for the fourth, fifth and sixth resonances. Accurate determination of width of these states are in the pipeline with more configurations in the basis set. To the best of our knowledge, this is the first prediction of the parameters of low lying resonances of heliumlike uranium ion and therefore warrants precise experimental verification.

\begin{table}[t]
\caption{\label{tab:resoure} Parameters of first six resonance states of heliumlike U ($Z=92$) below $\mathrm{U^{91+}}$($2s$) threshold.
}
\begin{ruledtabular}
\begin{tabular}{ccc}
      & Resonance energy & Resonance width\\
State &\multicolumn{1}{c}{$E_{r}$ (a.u.)}&$\Gamma$ (a.u.)\\ 
\hline
1 & --2497.1405	     &0.01682\\
2 & --2490.8784      &0.00126\\
3 & --2159.9825      &0.00096\\
4 & $\sim$ --1785.81 &$-$\\
5 & $\sim$ --1783.11 &$-$\\
6 & $\sim$ --1567.75 &$-$
\end{tabular}
\end{ruledtabular}
\end{table}

The present results with $Z=$ 2 and 92 establishes the applicability of the relativistic configuration-interaction (CI) framework to the  stabilization method for any two-electron system. More CSFs ($l>2$) in the present framework will certainly increase the possibility of locating more number of resonances below the threshold. It would also be interesting to see the effect of QED on the parameters of these resonances in future. In order to construct the resonance wavefunction, we can choose the $R$ value for an appropriate eigenvalue at which the DOS reaches its maximum. However, in case of very narrow resonances, $R$ value at the midpoint of the plateau of a particular eigenvalue can fairly be chosen. Having the idea of the resonance wavefunction more structural information about the state may be extracted.

This method is applicable to any total angular momentum state ($^{2s+1}$L$_J$) of heliumlike ions and in principle to the complex resonance structures of many-electron systems. It can also be used to study the behavior of bound states of few-electron ions inside a finite domain to realize the pressure confinement. As an example, we choose a two-electron ion embedded in dense plasma environment belonging to strong coupling regime considering the ion-sphere (IS) model potential \cite{ichimaru1982} where the two-body interaction between the nucleus and a bound electron is given by
\begin{eqnarray}\label{is}
V_{IS}(r_i)=-\frac{Z}{r_i}+\frac{Z-N_e}{2R}\left[3-\left(\frac{r_i}{R}\right)^2\right]
\end{eqnarray}
Here $Z$ is the nuclear charge of the positive ion and $N_e$ is the number of bound electrons. $R$ is the IS radius which is determined from the charge neutrality condition
\begin{eqnarray}
Z-N_e =\frac{4}{3}\pi R^3 n_e
\end{eqnarray}
The density of plasma electrons ($n_e$) thus governs the size of the sphere (Wigner-Seitz sphere) \cite{ichimaru1982}. Due to the spatial finiteness of the IS potential, the wavefunction of the bound electrons is truncated at the boundary of the Wigner-Seitz sphere. Considering the experimental interest on the dense aluminum plasma \cite{Vinko2015}, we have made an attempt to estimate the ground state energy of heliumlike aluminum ion as a test case. It has been found that the present method yields ground state energy eigenvalues --156.653 a.u., --151.299 a.u. and --148.682 a.u. for $n_e=$ 5$\times 10^{22}$ per cm$^3$, 5$\times 10^{23}$ per cm$^3$ and 1$\times 10^{24}$ per cm$^3$ respectively. The present values are in reasonable agreement with those estimated by Chen \textit{et. al.} \cite{chen2019} using multi-configuration Dirac-Fock (MCDF) wave functions with the inclusion of a finite nuclear size and QED corrections.

\section{Conclusion}\label{conc}
The stabilization method in the relativistic configuration interaction framework has been adopted to investigate the isolated  resonances of He and heliumlike uranium ion with $\Pi=0$ and $J=0$. The ions are considered to be enclosed within an impenetrable spherical cavity, the radius of which is varied continuously. The stabilization diagram is therefore realized with respect to the radius of the cavity ($R$). The advantage of this diagram is that one can obtain the tentative positions of all the resonances admissible for a given basis size. It has been noted that a single eigenroot of the diagonalized Hamiltonian may form a flat plateau in the vicinity of avoided crossings at several resonance energies. Therefore, different roots show flat plateau in the vicinity of avoided crossings for particular resonance energy. The DOS profiles are used to determine the parameters of the resonance states. In case of isolated resonances, the DOS is numerically estimated by taking the inverse of tangent at different points near the flat plateaus which are expected to fit with perfect lorentzian functions. But the shape of the numerically estimated DOS profile depends on two major factors: Firstly the quality and size of the basis set which actually determines whether a resonance is properly localized by a particular eigenroot of the diagonalized Hamiltonian or not. Therefore, a distorted DOS profile for a particular root implies that the root does not adequately represent the given resonance. We have shown the dependence of the DOS profiles corresponding to different resonances on the diagonalized eigenroots. Nevertheless, we can choose the best fitted DOS profile (looking at the statistical fitting parameters $\chi^2$ and $\mathcal{R}^2$) from different roots to estimate the position and width of a particular isolated resonance. The second important factor is the mesh size of the basis set/stabilizing parameters (here in the present work, it is the radial extent of the wavefunction $R$). Due to obvious reason, the accuracy of the numerical value of DOS at different energy positions in the neighbourhood of the plateau is quite sensitive on the mesh size. Lesser the mesh size, variation of tangent at different points in the plateau region will be more accurate. For narrow resonance or in case of two closely spaced isolated resonances (not necessarily narrow), the numerical errors can essentially be avoided by considering lesser mesh size of $R$. We have determined the parameters of two closely spaced resonances of helium-like uranium accurately just by increasing the mesh size. In summary, there is always a scope of improving the accuracy of the DOS profiles and therefore the precision of the estimated resonance parameters by increasing the number of terms in the basis set (or the radial extension of the basis set) as well as by decreasing the mesh size of the stabilizing parameter. In the strong confinement regions, we have also determined the variation of the bound state energies \textit{ w.r.t.} the cavity radius ($R$) and made an estimate about the pressure felt by the atom inside the cavity. Moreover, we have shown how the present method can be useful to determine the bound state energies of highly charged ions in strongly coupled plasma environment. This hybrid technique has the potential to open up a new horizon on investigating the resonance structures of highly charged ions of high-$Z$ elements.


%
\begin{acknowledgments}

This research was supported in part by Funda\c{c}\~{a}o para a Ci\^{e}ncia e a Tecnologia (FCT), Portugal, through the research center Grants No. UID/FIS/04559/2019 and No. UID/FIS/04559/2020 (LIBPhys), from FCT/MCTES/ PIDDAC, Portugal.
PA acknowledges the support of the FCT, under Contract No. SFRH/BPD/92329/2013.
JKS acknowledges the partial financial support from the DHESTBT, Govt. of West Bengal under grant number 249(Sanc.)/ST/P/S \& T/16G-26/2017. The financial assistance provided through Grant No. 23(Sanc.)/ST/P/S \& T/16G-35/2017 by DHESTBT, Govt. of West Bengal, India is acknowledged by SB.

\end{acknowledgments}
\RaggedRight
\bibliography{biblio}


\end{document}